\documentclass[12pt,preprint]{aastex}

\slugcomment{
Asai: 2011/12/25
}

\shorttitle{Moreton Wave and EIT Wave}
\shortauthors{Asai et al.}

\begin{document}

\title{First Simultaneous Observation of H$\alpha$ Moreton Wave, EUV
Wave, and Filament/Prominence Oscillations}

\author{
Ayumi Asai\altaffilmark{1},
Takako T. Ishii\altaffilmark{2},
Hiroaki Isobe\altaffilmark{1},
Reizaburo Kitai\altaffilmark{2},
Kiyoshi Ichimoto\altaffilmark{2},
Satoru UeNo\altaffilmark{2},
Shin'ichi Nagata\altaffilmark{2},
Satoshi Morita\altaffilmark{2},
Keisuke Nishida\altaffilmark{2},
Daikou Shiota\altaffilmark{3},
Akihito Oi\altaffilmark{4},
Maki Akioka\altaffilmark{5},
and
Kazunari Shibata\altaffilmark{2}}

\email{asai@kwasan.kyoto-u.ac.jp}
\altaffiltext{1}{
Unit of Synergetic Studies for Space, Kyoto University,
Yamashina, Kyoto 607-8471, Japan.}
\altaffiltext{2}{
Kwasan and Hida Observatories, Kyoto University,
Yamashina, Kyoto 607-8471, Japan.}
\altaffiltext{3}{
Advanced Science Institute, RIKEN, Wako, Saitama 351-0198, Japan.}
\altaffiltext{4}{
College of Science, Ibaraki University, Mito, Ibaraki 310-8512, Japan.}
\altaffiltext{5}{
Hiraiso Solar Observatory, National Institute of Information and
Communications Technology, Hitachinaka, Ibaraki 311-1202, Japan.}

\begin{abstract}
We report on the first simultaneous observation of an H$\alpha$ Moreton
wave, the corresponding EUV fast coronal waves, and a slow and bright
EUV wave (typical EIT wave).
Associated with an X6.9 flare that occurred on 2011 August 9 at the
active region NOAA 11263, we observed a Moreton wave in the H$\alpha$
images taken by the Solar Magnetic Activity Research Telescope (SMART)
at Hida Observatory of Kyoto University.
In the EUV images obtained by the Atmospheric Imaging Assembly (AIA) on
board the {\it Solar Dynamic Observatory} ({\it SDO}) we found not only
the corresponding EUV fast ``bright'' coronal wave, but also the EUV
fast ``faint'' wave that is not associated with the H$\alpha$ Moreton
wave.
We also found a slow EUV wave, which corresponds to a typical EIT wave.
Furthermore, we observed, for the first time, the oscillations of a
prominence and a filament, simultaneously, both in the H$\alpha$ and EUV
images.
To trigger the oscillations by the flare-associated coronal disturbance,
we expect a coronal wave as fast as the fast-mode MHD wave with the
velocity of about 570 -- 800~km~s~$^{-1}$.
These velocities are consistent with those of the observed Moreton wave
and the EUV fast coronal wave.
\end{abstract}


\section{Introduction}
Moreton waves, flare-associated waves seen in H$\alpha$, have been
observed \citep{mor1960,smi1971,shi2011} to propagate in restricted
angles with the velocity of about 500 -- 1500~km~s$^{-1}$.
They sometimes show arc-shaped fronts, and are often associated with
type-II radio bursts \citep{kai1970}.
They are transient, and appear only for about 10 minutes.
Associated with flares, remote filaments and prominence are sometimes
activated or excited to oscillate.
These ``winking filaments'' are also thought to be caused by
flare-associated waves, and are called as invisible Moreton waves
\citep{smi1971,tri2009,her2011}.
After the findings, \citet{uchi1968} suggested that Moreton waves are
the intersection of the fast-mode magnetohydrodynamic (MHD) shock
propagating in the corona with chromosphere.
This model has been widely accepted, and the coronal counterparts have
been surveyed for a few decades.
Moreton waves are rare to be observed even for large flares \citep{shi2011}.

After the launch of the {\it Solar and Heliospheric Observatory} ({\it
SOHO}), the EUV Imaging Telescope (EIT) found wavelike phenomena
associated with flares, which are called ``EIT waves''
\citep{thom1999,thom2000}.
Although EIT waves were expected to be the coronal counterpart of
Moreton waves, they show different physical characteristics from those
of Moreton waves:
the propagating velocity is much slower than that of Moreton wave and is
about 200 -- 400~km~s$^{-1}$, the lifetime is much longer and is about
45 -- 60 minutes, they can show isotropic propagation, while Moreton
waves propagate with restricted angles \citep{kla2000,war2007,thom2009}.
There have been, therefore, remained a question whether EIT waves are
really coronal counterparts of Moreton waves or no.
As for searching for a coronal counterpart of Moreton waves, the Soft
X-ray Telescope (SXT) on board {\it Yohkoh} found wavelike phenomena in
soft X-rays, called X-ray waves \citep{khan2000,khan2002}.
X-ray waves are confirmed to be a real counterpart of Moreton waves by
simultaneous observations of X-ray waves and Moreton waves
\citep{naru2002,naru2004}.

Then, we come to an issue what EIT waves are.
\citet{eto2002} clearly showed that an EIT wave is different from a
Moreton wave, based on simultaneous observations of them.
On the other hand, \citet{war2004a,war2004b} argue that the velocity
discrepancy of EIT and Moreton waves can be explained by the
deceleration of coronal waves.
The mechanism of EIT waves remains, therefore, very controversial
\citep{war2007,will2009,gall2010}.
\citet{del1999} and \citet{chen2002,chen2005} proposed the field-line
stretching model for EIT waves.
They suggested that EIT bright fronts were not ``waves'' at all, but
instead plasma compression at stable flux boundaries due to rapid
magnetic field expansion.
This model can also resolve the puzzle why EIT waves often stop at
magnetic separatrices.

Recently, by the Atmospheric Imaging Assembly (AIA; Title \& AIA team
2006, Lemen et al. 2011) on board the {\it Solar Dynamic Observatory}
({\it SDO}), fast coronal waves have been observed associated with flares 
\citep[e.g.][]{liu2010,ma2011}.
These waves (hereafter called ``EUV fast coronal waves'') are thought to
be the fast-mode MHD waves.
Coronal X-ray waves and EUV fast coronal waves have been also observed
spectroscopically with the EUV Imaging Spectrometer (EIS) on board {\it
Hinode} \citep{asa2008,harra2011}.

\citet{chen2011} found two different coexisting coronal waves, slow
coronal wave (i.e. EIT wave) and fast coronal wave, from EUV
observations taken by {\it SDO}/AIA.
Although the fast coronal wave seems to be the coronal counterpart of a
Moreton wave, they used only EUV images, and it remained to be confirmed
whether it is a classical H$\alpha$ Moreton wave.
This letter presents the first simultaneous observation of EUV waves and
a Moreton wave by using EUV and H$\alpha$ images with high spatial and
temporal resolutions.
Moreover, we found not only a winking filament on the disk, but also an
oscillating prominence on the limb, triggered by the coronal wave
(Moreton wave).

\section{Observations}
\begin{figure}
\epsscale{.90}
\plotone{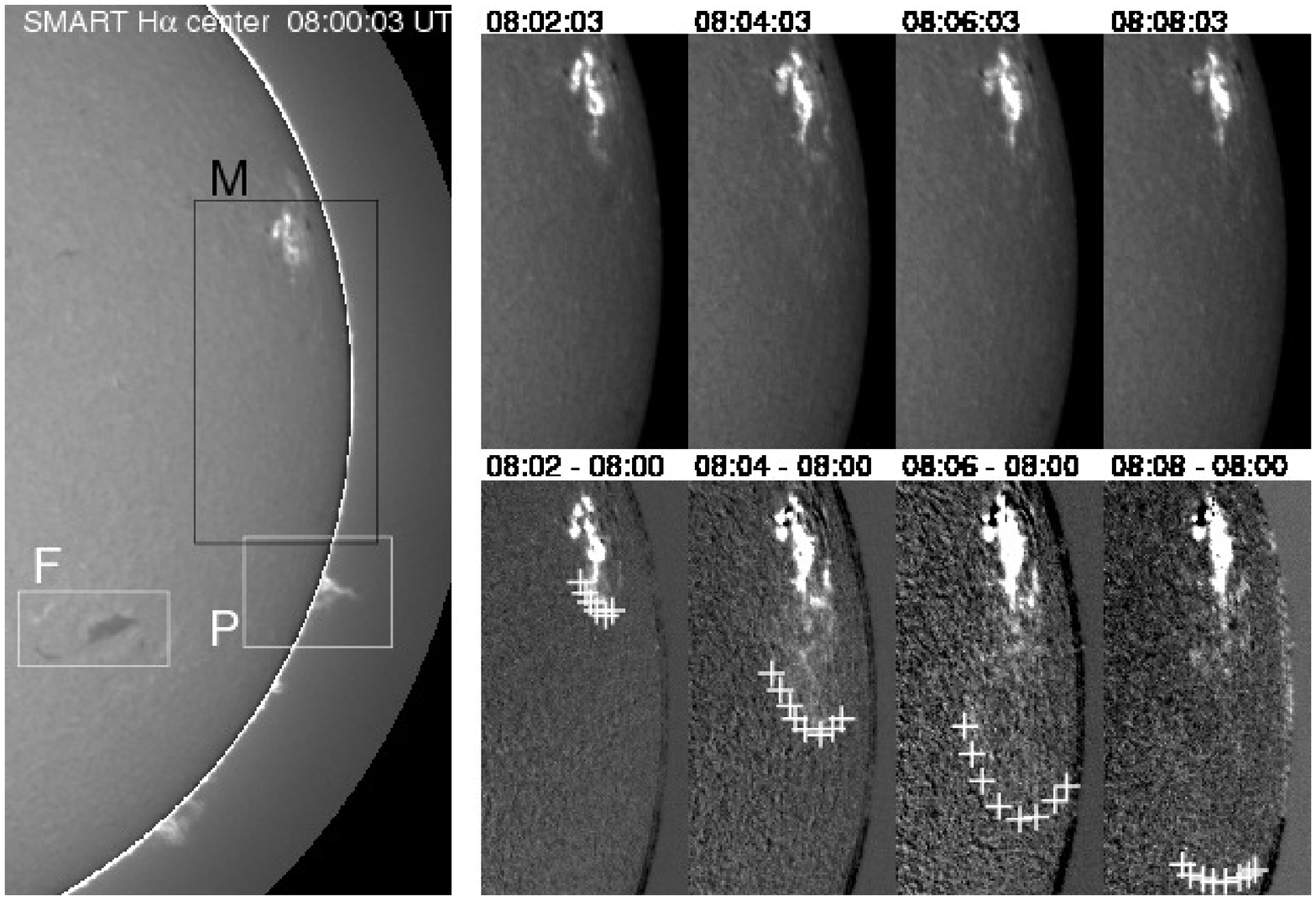}
\caption{Propagation of the Moreton wave observed with SMART.
Solar north is up and west to the right.
The top right panels are the sequence of H$\alpha$ images of the region
``M'' (the black rectangle) in the left panel.
The bottom right panels are the same sequence of the H$\alpha$ images
but are subtracted with the image at 08:00~UT.
The white rectangles with marks ``P'' and ``F'' in the left panel
indicate the oscillated prominence and filament, respectively.
\label{moreton}}
\end{figure}

An intense flare, X6.9 on the {\it GOES} scale, occurred on 2011 August
9 at the Active Region NOAA 11263 (N17$^{\circ}$, W71$^{\circ}$).
The flare started at 07:48~UT, and peaked at 08:05~UT.
Associated with the flare, we observed a Moreton wave in the H$\alpha$
images obtained by the Solar Magnetic Activity Research Telescope
(SMART; UeNo et al. 2004), at Hida Observatory, Kyoto University, Japan.
SMART regularly observes the full-disk sun in seven wavelengths around
the H$\alpha$ line (6562.8~{\AA}), i.e., H$\alpha$ center and the wings
at $\pm$0.5, $\pm$0.8, and $\pm$1.2~{\AA}.
The time cadence is 2 minutes for each wavelength during the impulsive
phase of the flare, and the pixel size is 0.56$^{\prime\prime}$.
Such full-disk and multi-wavelength observation with high cadence is
suitable to detect Moreton waves \citep[e.g.][]{naru2008}.
The Moreton wave was seen only from 08:02~UT to 08:08~UT with the SMART
data.
Figure~\ref{moreton} shows the Moreton wave in H$\alpha$ center images
taken by SMART.
It mainly traveled in the south direction from the flare site.
We derived the propagation speed by following the fastest wavefront.
The mean propagation speed during the 6-minute appearance of the Moreton
wave was about 760~km~s$^{-1}$.
The H$\alpha$ $-1.2$~{\AA} images taken with SMART show the ejection of
a filament with the velocity of about 300~km~s$^{-1}$ in the direction
of the Moreton wave\footnote{see,
http://www.kwasan.kyoto-u.ac.jp/topics/110809/bin\_p12/}.

To compare the physical features of H$\alpha$ Moreton waves with
wave-like phenomena observed in EUVs, we used EUV images taken by {\it
SDO}/AIA.
In this letter we mainly used the 193~{\AA} images, which are mainly
attributed to the Fe~{\sc xii} (log($T$) $\sim$ 6.1) line.
The temporal resolution of the AIA 193~{\AA} data during the flare was
12 second.
The flare site was close to the west limb.
We also used EUV images taken by the Extreme-Ultraviolet Imager (EUVI)
of the Sun Earth Connection Corona and Heliospheric Investigation
(SECCHI; Howard et al. 2008) on board the {\it Solar Terrestrial
Relations Observatory} ({\it STEREO}; Kaiser et al. 2008){\it -Ahead}
satellite ({\it STEREO-A}).
The {\it STEREO-A} was $\sim$100.7$^{\circ}$ ahead of the earth at the
time of the flare.
The temporal resolutions of the 195 and 304~{\AA} data, which we used in
this letter, were 5 and 10 minutes, respectively.
The pixel size of the images is 1.58$^{\prime\prime}$.

Figure~\ref{euv} shows the wave propagation seen in the {\it SDO}/AIA
193~{\AA} images ({\it top}) and in the {\it STEREO-A}/EUVI 195~{\AA}
images ({\it bottom}).
All are difference images, and are subtracted by the intensity maps
taken at 07:55:19 and 07:55:31~UT for AIA and EUV data, respectively.
The right panels of Figure~\ref{euv} show the potential magnetic field
lines for the view from the earth ({\it top}) and from the {\it
STEREO-A} ({\it bottom}).
The potential magnetic field lines are calculated by using synoptic
magnetograms from GONG data\footnote{http://gong.nso.edu/data/magmap/}
and based on the method by \citet{shio2008}.
In the images at 08:05~UT we can identify sharp wave fronts traveling
southward, like as reported by \citet{thom2000}, while the sharp fronts
disappear after 08:10~UT.
The EIT wave was blocked by small ARs in traveling, and did not show the
isotropic feature.
The magnetic field of the southern region of the flare site is weak.
A part of the EUV wave traveling southward, which is shown with the
arrows in Figure~\ref{euv}, traveled without being disturbed by such
ARs.
The propagation velocity of about 700~km~s$^{-1}$ was measured by
following the wavefronts in the images.
The wave front is much fainter than the sharp front seen at 08:05~UT.

\section{Analysis and Results}
In Figure~\ref{shock}(a) and (b) we showed the comparison of the spatial
structure of the Moreton wave with that of the EUV wave of this flare.
These are the difference images in AIA 193~{\AA} and in SMART H$\alpha$
center, respectively.
The time difference is noted in the Figure.
The plus ($+$) signs follow the front of the Moreton wave.
The front is well coincident with the sharp bright EUV wave front.
In addition, we can identify an expanding dome on the wave front in AIA
EUV images.
Therefore, the expanding dome is thought to be the shock traveling in
the corona, and the Moreton wave and the sharp EUV wave are the
intersection with the chromosphere.

We examined the temporal features of the EUV wave, by using the
time-distance diagrams (time-slice images) along the lines shown in the
Figure~\ref{shock}(a).
Figure~\ref{shock}(c), (d), and (e) are the time-slice images for the
Line 1, 2, and 3, respectively.
For each diagram, the flare site was set to be zero.
The velocities were derived by following the features in the time-slice
diagrams.
The Line 1, which is drawn with a straight line mainly follows the dome
structure expanding in the corona.
In the time-slice images (Figure~\ref{shock}c) we can see a bright front
(F1) with the traveling velocity of about 760~km~s$^{-1}$, while it is
initially even faster.
Behind the front, we can identify the dimming feature.
The Line 2 mainly follows the same path of the Moreton wave, and is
following a great circle of the solar surface from the flare site.
The front is very bright and sharp from 08:01 to 08:09~UT (F2b), which
is almost the same time range of the Moreton wave.
Even after 08:09~UT, we can identify the wave front, while it became
much fainter (F2f).
The traveling velocities are about 730 and 620~km~s$^{-1}$ for F2b and
F2f, respectively.
On the other hand, the bright edge is rapidly decelerated (S2b) and
disappears at 08:12~UT.
We also draw the Line 3 that also follows a great circle of the solar
surface.
Although the direction of the Line 3 is out of the arc of the Moreton
wave front, we can see a wave-like feature.
The propagation velocity of the bright front is initially about
550~km~s$^{-1}$ (F3b), and slow down to be about 340~km~s$^{-1}$ (S3b)
after 08:06~UT.
The slow bright EUV wave (S3b) suddenly disappears at about 08:12~UT.
On the other hand, we can identify a fast faint feature (F3f) from
08:06~UT.  The velocity is about 580~km~s$^{-1}$.
As we mentioned above, there are small ARs on the passes of the EUV
waves.
These small closed loops start to oscillate due to the propagation of
the coronal wave.
Along the Line 2 and 3, we can identify some oscillations as shown with
the white arrows in Figure~\ref{shock}(d) and (e).

\begin{figure}
\epsscale{0.5}
\plotone{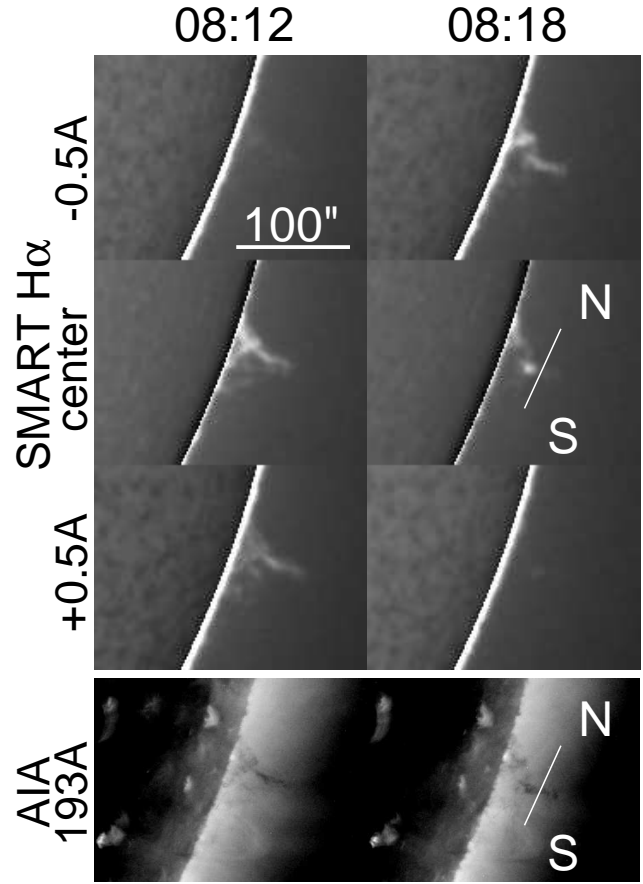}
\caption{Temporal evolution of the oscillating prominence.
{\it From top to bottom}: H$\alpha$ $-0.5$~{\AA}, center images, and
$+0.5$~{\AA} images taken by SMART, and EUV 193~{\AA} images taken by
{\it SDO}/AIA.
The field of view of each panels is the region P shown in
Figure~\ref{moreton}.
The times (UT) are indicated on the top.
\label{osci}}
\end{figure}

Associated with the flare, we observed the oscillations of a prominence
on the west limb and a filament on the disk.
The sites of the prominence/filament are shown in the left panel of
Figure~\ref{moreton}, with the characters P (prominence) and F
(filament), respectively.
In Figure~\ref{osci} we show the temporal evolution of the limb
prominence (P) in H$\alpha$s and in EUV (193~{\AA}) taken by SMART and
AIA.
From the SMART H$\alpha$ wing data, we can clearly see the prominence
moving in the line of the sight.
First, the prominence became bright in the plus wing.
The initial sign of the oscillation is identified with the darkening of
the prominence in the $-0.5$~{\AA} image taken at 08:11:57~UT.
In the sequence of 08:14~UT, (at least, a part of) the prominence is the
brightest in the most redward wing image of the observation,
i.e. $+1.2$~{\AA} image (08:14:24~UT).
This means that the prominence is moving away with Doppler velocity of
about 50~km~s$^{-1}$ or even faster.
After this, the motion of the prominence turned to be blueward, and in
the sequence of 08:22~UT, it is brightest in the $-1.2$~{\AA} images
(08:21:45~UT).
The oscillation period is roughly 15 minutes.
In the AIA EUV images, on the other hand, we can also see the motion in
the plane.
First, it moves southward (downward in the images), then moves northward
(upward) after 08:17~UT.
The oscillation period is about 12 -- 16 minutes, and the amplitude is
about 10,000~km.
The apparent velocity is about 30~km~s$^{-1}$.
In Figure~\ref{hei_tim}(a) we show the time-slice image of the
oscillating prominence in EUV (193~{\AA}) overlaid with the brightest
points of the prominence in the H$\alpha$ center images (the plot of the
$+$ sings).
The slit position is shown in Figure~\ref{osci}.
The prominence oscillation is identified as a filament oscillation in
the {\it STEREO-A}/EUVI 304~{\AA} images, and we used the images to
determine the precise distance from the flare site.
From the start time of the oscillation and the site of the prominence,
the coronal wave propagating with the velocity of about 800~km~s$^{-1}$
is expected to activate the prominence.

The filament on the disk F also showed oscillation features, although
the features are much weaker both in the line-of-the-sight direction and
in the plane than those for the prominence P.
The start of the oscillation is roughly estimated as 08:17:57~UT from
the H$\alpha$ wing images.
The oscillation period is about 15 minutes.
We have to note that we can identify small activation at the footpoints
of the filament F at 08:01:31~UT.
Therefore, some weak movements of the filament already started when the
coronal disturbance arrived at the filament.
From the start time of the oscillation and the site of the filament, we
expect the coronal wave propagating with the velocity of about
570~km~s$^{-1}$.

\section{Summary and Conclusions}
\begin{figure}
\epsscale{1.}
\plotone{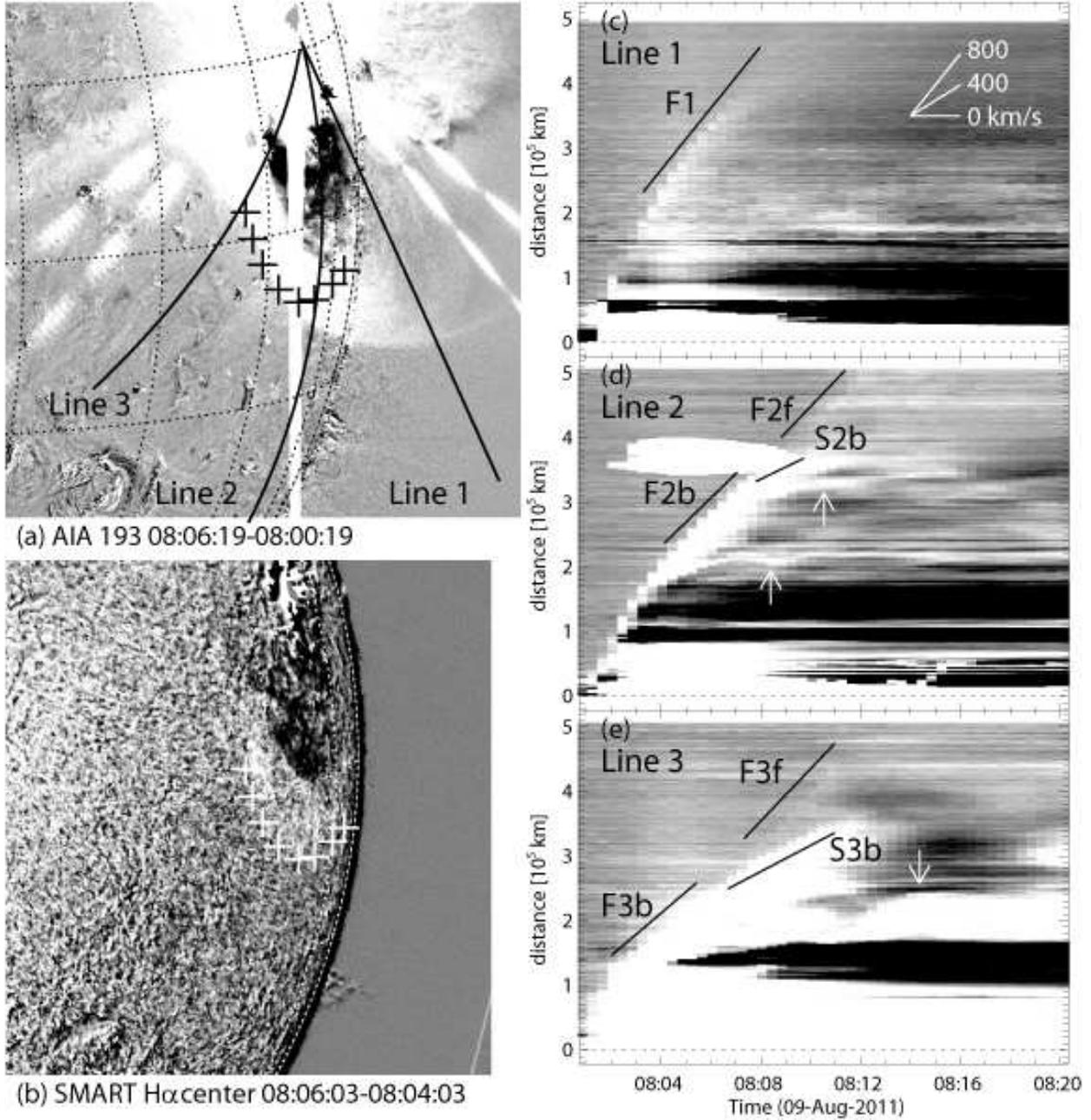}
\caption{Detailed feature of coronal disturbances.
(a) A difference image (08:06:19$-$08:00:19~UT) of EUV 193~{\AA} image
taken by {\it SDO}/AIA, and
(b) a difference image (08:06:03$-$08:04:03~UT) of H$\alpha$ center
image taken by SMART.
The plus ($+$) signs follow the Moreton wave front.
(c $\sim$ e) Time-distance diagrams (time-slice images) of the AIA EUV
193~{\AA} image along the line 1, 2, and 3, respectively.
The lines are shown in (a).
The solid lines marked with F1, F2b, F2f, F3b, F3f, and S3b follow EUV
wave fronts: F (Fast) or S (Slow), the number of the line, and b
(bright) or f (faint).
The white arrows point the oscillating features caused during the
propagation of the EUV waves.
\label{shock}}
\end{figure}

\begin{figure}
\epsscale{0.7}
\plotone{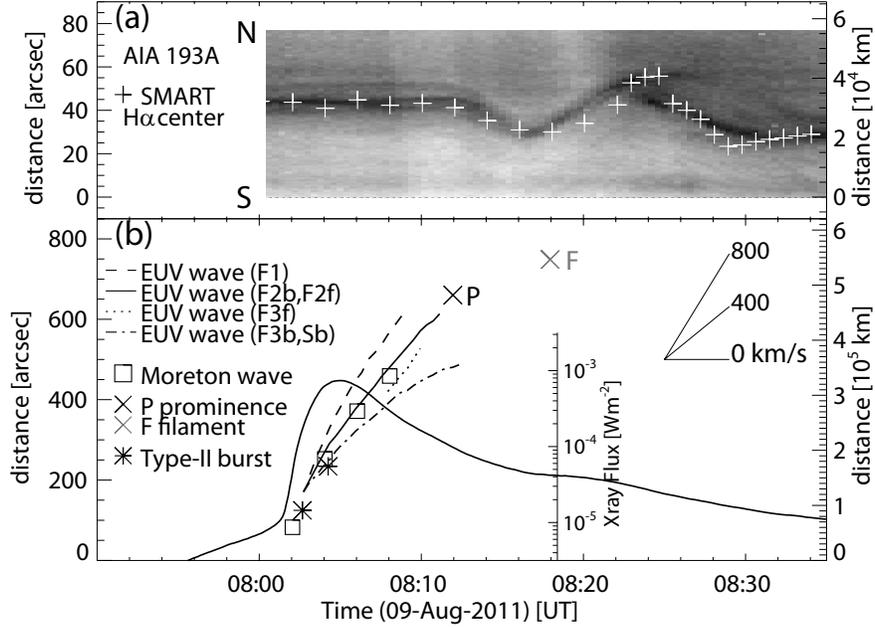}
\caption{Height-time evolution of the flare-related phenomena.
(a) A time-distance diagram (time-slice image) of the AIA EUV 193~{\AA}
image along the slit line shown in the Figure~\ref{osci}.
N and S show the north and south directions, respectively.
The brightest points of the prominence in the H$\alpha$ center images
are overlaid with the plus ($+$) signs.
(b) Temporal evolutions of the EUV waves, the Moreton wave, and the
type-II radio burst with GOES X-ray flux at the 1 -- 8~{\AA} channel.
The distance of the Moreton wave front ($\Box$) was measured along a
great circle of the solar surface from the flare site.
The EUV wave fronts measured with the Line 1, 2, and 3 (see,
Figure~\ref{shock}) are shown with the dashed, solid, and dash-dot
lines, respectively.
The fast faint EUV wave seen in the Line 3 (F3f) is also shown with the
dotted line.
The times and distances of the oscillating limb prominence (P) and disk
filament (F) are shown with the cross ($\times$) signs.
The distance of the type-II radio burst ($\ast$) was calculated, by
using a coronal electron density model, and the zero was set to be the
photosphere.
\label{hei_tim}}
\end{figure}

We simultaneously observed the H$\alpha$ Moreton wave and the
corresponding EUV fast coronal wave.
The Moreton wave front was well consistent with the fast-bright-sharp
EUV wave front (F2b).
Even after the Moreton wave disappeared, we identified the propagation
of the fast-faint EUV wave (F2f).
Even along the Line 3 (Figure~\ref{shock}a), which is the direction
without the Moreton wave, we found that the fast EUV waves (F3b, F3f).
The fast EUV waves (F1, F2f, F3b, and F3f) are thought to be the
fast-mode MHD waves (coronal waves).
The EUV fast coronal waves appear more frequently than Moreton wave,
although they are very faint and have not been observed until the launch
of {\it SDO}.
Especially, only when the shock strongly contacts with the chromosphere,
the intersection is observed as the Moreton wave (F2b).
The temporal evolutions of the H$\alpha$ Moreton wave and the EUV waves
are summarized in Figure~\ref{hei_tim}(b).

In Figure~\ref{hei_tim} we also show the temporal evolutions of a
type-II radio burst associated with the flare.
In the metric radio spectrogram (25 -- 2500~MHz) observed with the
Hiraiso Radio Spectrograph (HiRAS; Kondo et al. 1995), we identify the
type-II radio burst from 08:02:40 to 08:06:30~UT.
Assuming the coronal density model proposed by \citet{new1961} and
\citet{mann1999}, we derived the propagation velocity of about
850~km~s$^{-1}$.
The observed type-II radio burst seems to be consistent with the Moreton
wave and fast bright EUV wave, which also supports the interpretation of
the wave as a fast-mode MHD shock.

We also found oscillations of a prominence and a filament.
To trigger the oscillations by a flare-associated coronal disturbance,
we expect a coronal wave as fast as the fast-mode MHD wave with the
velocity of about 570 -- 800 km~s~$^{-1}$.
These velocities are consistent with the propagation velocities of the
observed Moreton wave and the EUV fast coronal wave.
An invisible Moreton wave could be such an EUV fast-faint coronal wave.
It is known that a typical slow EIT wave sometimes causes filament
oscillations \citep{oka2004}, or such filament oscillations triggered by
an EIT wave are expected to be stronger than those by an invisible
Moreton wave (P. F. Chen, private communication).
In the current case, however, the role of the EIT wave on the
filament/prominence oscillations is unclear.

\begin{figure}
\epsscale{1.}
\plotone{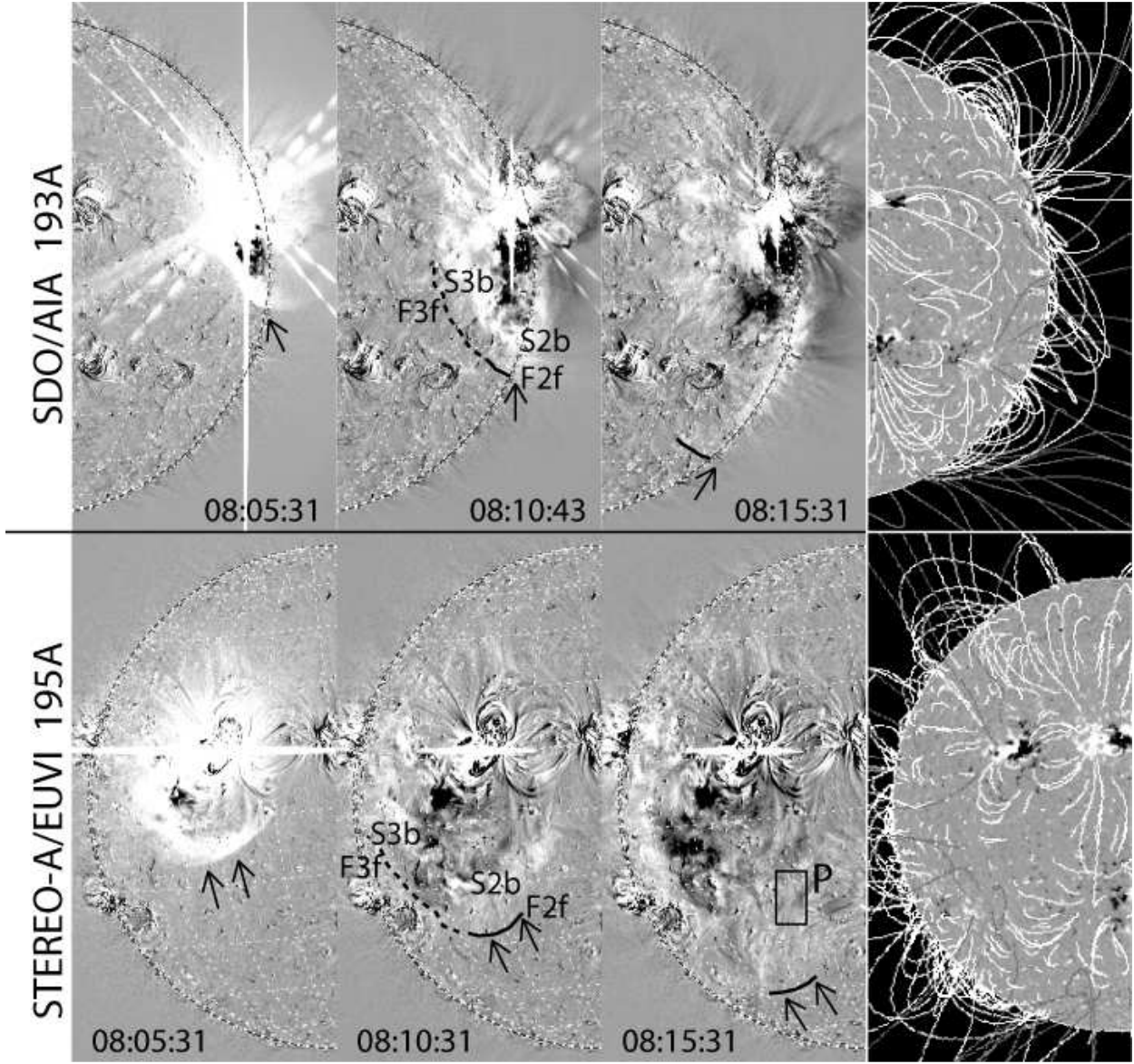}
\caption{EUV waves observed with {\it SDO}/AIA 193~{\AA} ({\it top}) and
{\it STEREO-A}/SECCHI/EUVI 195~{\AA} ({\it bottom}).
The arrows follow the front of the EUV fast coronal wave.
The region with mark P points the position of the limb prominence.
The right panels show the potential magnetic field lines for the views
from the earth ({\it top right}) and from the {\it STEREO-A} ({\it
bottom right}).
\label{euv}}
\end{figure}

Along the Line 2 and 3, we identified slow-bright EUV waves (S2b, S3b)
behind the fast-faint EUV waves.
From the propagating features (i.e. the velocity and the isotropic
propagation; Figure~\ref{euv}), we think it is a typical EIT wave.
We simultaneously observed the EIT wave and the EUV fast wave (fast-mode
MHD wave) as reported by \citet{chen2011}.
This means that the EIT wave is different from a fast-mode MHD wave,
which supports the field-line stretching model
\citep{chen2002,chen2005}.
It is, however, difficult to clearly identify both features separately
in the very initial phase of the flare, and it is unclear the relation
between the fast-bright (F3f) and the slow-bright (S3b) EUV waves along
the Line 3.

There is an alternative possibility that we observed a single coronal
disturbance, and the two different waves (F2f/F3f and S2b/S3b)
correspond to the fronts of two different heights (faster ones are
higher) of the disturbance \citep{vero2008,war2011}.

We, however, do not think the possibility due to the following reasons:
First, the slow bright waves (S2b, S3b) stopped propagating at small
active regions.
This conflicts with the features of a fast-mode wave, 
while this is possibly reconsiled by considering the stopping front as 
CME flanks as reported by \cite{pat2009}.
Second, we observed the prominence/filament oscillations.
Since they are located low in the corona, we can derive the velocity of
the shocks/waves there, from the distances and the times of the
oscillations.
In the current case we need waves with velocities of 570~km~s$^{-1}$ or
more.
Especially, the direction of the filament is close to the Line 3, and
the required velocity is much faster than the bright slow wave (S3b),
while it is consistent with the fast faint wave (F3f).

\acknowledgments

We thank the referee for the useful comments.
This work is supported by KAKENHI (23340045),
and by the Global COE Program ``The Next Generation of Physics, Spun
from Universality and Emergence'' from MEXT, Japan.



\clearpage

\end{document}